# Preparation and analysis of strain-free uranium surfaces for electron and x-ray diffraction analysis

*JE Sutcliffe[1], JR Petherbridge[2], T Cartwright[2], R Springell[1], TB Scott[1]*

*JE Darnbrough[1]*

[1]*Interface Analysis Centre, HH Wills Physics Laboratory, University of Bristol, Tyndall Avenue, Bristol, BS8 1TL, UK*

[2]*AWE Plc, Aldermaston, Reading, Berkshire, RG7 4PR, UK*

*13 February 2018*

## Abstract
This work describes a methodology for producing high quality metallic surfaces from uranium primarily for characterisation and investigations involving electron backscatter diffraction. Electrochemical measurements have been conducted to inform ideal polishing conditions to produce surfaces free from strain, induced by mechanical polishing. A commonly used solution for the electropolishing of uranium, consisting in part of phosphoric acid, was used to conduct the electrochemical experiments and polishing. X-ray diffraction techniques focusing on the surface show low stresses and strains are exhibited within the material. This is mirrored in good quality electron backscatter diffraction.

## Introduction
Strong scattering of x-rays and electrons in atomically heavy (high-Z) materials greatly inhibits penetration and results in sampling of the near-surface region [1]. Therefore, excellent preparation of high-Z materials such as uranium is paramount for conventional, diffraction-based techniques. The susceptibility of the surface to mechanical deformation and chemical attack during preparation, leads to changes in chemistry and microstructural defects inhibiting coherent diffraction [2]. For the surface to be representative of the bulk, it is essential that that artefacts arising from oxidation, work hardening, and preferential etching are removed.

Sequential mechanical and electrochemical polishing of uranium and its alloys is standard practice [3,4]. However, previous studies on uranium lack insight into the changes caused by this preparation method and how to perfect it. To develop and optimise the process, a consideration of the electrochemistry of the reaction being driven during electropolishing is required. Linear sweep and cyclic voltammetry is used to understand the kinetics of electron transfer and mass transport. Chronoamperometry (potential step voltammetry) was used to assess the diffusional characteristics of the setup and conduct polishing. Electrochemical testing was ultimately carried out inform appropriate polishing potentials and durations.

This paper looks to elucidate the process of producing strain- and oxide-free surfaces as characterised by x-ray diffraction (XRD) and electron backscatter diffraction (EBSD) in uranium metal. Strain-free surfaces allow interrogation of the bulk characteristics and are of interest for the assessment of microstructure, crystallographic texturing and corrosion properties.

## Materials and techniques

### Materials

Specimens used for this investigation were cast and rolled low carbon samples (~ 50 ppm) of depleted uranium (size: 10 x 10 x 0.75 mm) provided by AWE Plc. Specimens were subsequently cut to 3 x 10 x 0.75 mm to limit the number of original pieces used, ensuring homogeneity between samples.

### Mechanical preparation

Samples were initially polished using SiC P360 grit paper to remove gross oxide before being potted in Clarocit$^{TM}$ epoxy resin purchased from Buehler. Samples were subsequently polished with progressively finer papers from P180 through to P4000. This produces a topologically rough sample surface on the scale of electron microscopy, incapable of producing EBSD patterns.

Samples were ultrasonically cleaned in acetone and broken out of the resin. Samples were then rinsed in water, acetone and methanol to remove any grease and allowed to dry through evaporation. As specimens were prepared in a batch to increase uniformity, samples were subsequently placed under vacuum to preserve the surface before electropolishing.

### Electrochemistry Configuration

Literature shows a bias for phosphoric acid-based solutions for uranium electropolishing with empirical evidence provided to prove their suitability [1,3,4,5,6]. For this work, a commonly used solution of 46% ethanol, 27% ethylene glycol and 27% phosphoric acid (85% assay) by volume [7], was selected to produce a strongly acidic electrolyte with a minimal presence of water or oxygen which would lead to anodisation of polished samples.

Electropolishing was carried out in a three-electrode cell, Figure 1, utilising a Ag/AgCl reference electrode, a Pt wire as the counter electrode, and the sample as the working electrode. A Gamry 1000 potentiostat was used to conduct linear sweep, cyclic voltammetry and chronoamperometry experiments. Electropolishing was conducted under chronoamperometry conditions.

A stirring rate of 900 rpm was achieved using a Fisher Scientific stirrer and a magnetic flea. All values of potential are quoted relative to the Ag/AgCl reference (+ 0.230 V vs the standard hydrogen electrode). Values of current have been normalised by each sample's exposed area to the solution to give a current density in mA/cm$^2$.

After electropolishing, all samples were rinsed with acetone, methanol and wiped using methanol-moistened blue-roll to remove any residue.

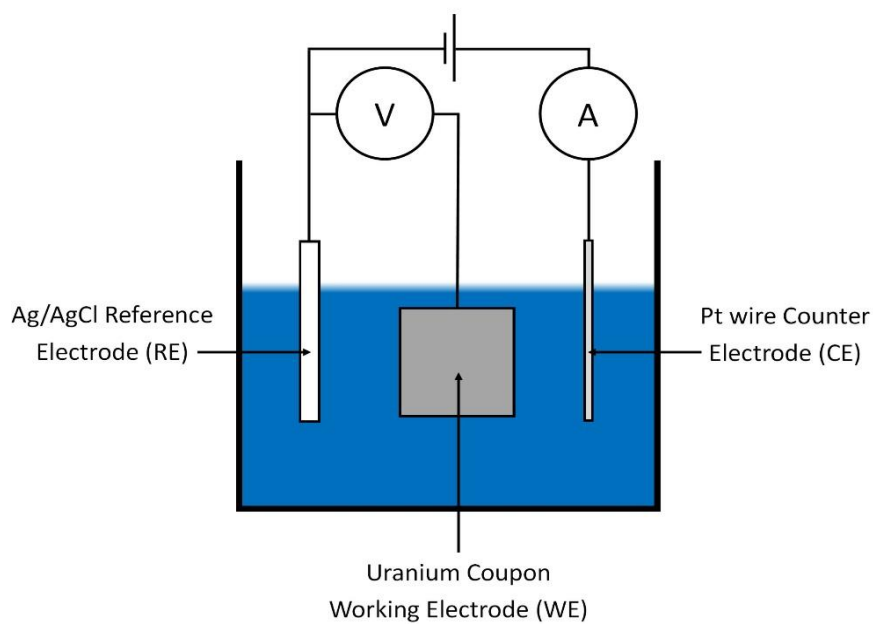

*Figure 1: Schematic of the experimental setup depicting the potentiostat acting as voltmeter, ammeter and supply.*

## Electron Backscatter Diffraction

SEM and EBSD was conducted on a Zeiss EVO MA10 scanning electron microscope fitted with a LaB$_6$ source, using a 20 µm aperture. A Digiview 3 high speed camera and EDAX OIM$^{TM}$ software were used to record and process EBSD data. Confidence indices were chosen as the predominant method of assessing EBSD pattern quality, due to the ability to easily compare datasets. Therefore, microscope and EBSD image processing settings were held fixed; 2 frames with an exposure time of 0.05 s were averaged for each scan point. All EBSD maps were run over areas sizing 100 µm x 50 µm. Image processing consisting of; background subtraction, dynamic background subtraction, normalised intensity histogram and median smoothing filter functions. No clean-up routines have been applied to any maps featured.

## X-ray Diffraction

XRD scans were obtained using a Philips X'Pert Pro multipurpose diffractometer utilising an Empyrean Cu K$_\alpha$ radiation source. The x-ray tube was operated at 40 kV and 40 mA with 0.04 radian Soller slits and 1/2° divergence slits. High angle powder scans were performed over a range of 25 to 140°.

Stress and strain within the sample were determined by two complementary XRD techniques. Firstly, Williamson-Hall peak analysis was applied to the powder data to assess strain and coherent volume size [8]. This method considers the change in peak width as the scattering vector is extended into the material normal to the surface.

The sin$^2\psi$ method assessed residual stresses by measuring the same Bragg peak at different effective sample tilts, thus sweeping the (fixed length) scattering vector progressively closer to the surface. A complete description of the technique is provided by Welzel at al. [9]. The effective sampling depth, $\tau_\omega$, is given by the following equation,

$$\tau_\omega = \frac{sin^2\theta - sin^2\psi}{2\mu \, sin\theta \, cos\psi}$$

(1)

θ is half the diffraction angle (2θ), ψ is the angle between the original surface normal and the resultant surface normal following sample 'tilting', and μ is the absorption coefficient for x-rays. For an 8 keV Cu K$_\alpha$ source, the absorption coefficient in uranium is equal to 5760 cm$^{-1}$ [10].

The (135) reflection, measured at a 2θ value of 131.5°, allowed for the greatest accessible value of sin$^2$ψ (0.8). Therefore, using the previous formula, a minimum information depth of 67 nm was achieved. Since the *d* spacing of this peak is 0.845 Å, fewer than 800 planes were accessed in this configuration. In comparison, the specular configuration, i.e. sin$^2$ψ=0, corresponds to greater than 9000 layers. Scans were performed over a range of 4° in 2θ with a step size of 0.02° and a counting time of 4 seconds per step. Positive tilts were used to avoid the beam footprint exceeding the size of the sample at high values of sin$^2$ψ.

For both analyses, peak fitting was undertaken using a least-squares fitting routine using the MATLAB M*fit* package produced at the Institut Laue-Langevin [11]. Peaks were fitted with Pseudo-Voigt functions using doublet peaks with wavelengths of 1.54059 and 1.54432 Å. Peaks were modelled with the Kα$_1$ line exhibiting twice the intensity of the Kα$_2$ line.

Full powder pattern analysis (including Rietveld refinement) was performed using GSAS-II to determine the lattice parameters and assess the texture of the material [12]. Due to significant preferred orientation caused by extensive processing, lattice and structural parameters used in the generation of material files for EBSD analysis were obtained from full pattern XRD fitting using GSAS-II.

# Results

## I-V Curve Characteristics

Figure 2 shows a typical single linear sweep starting in the cathodic region (-2 V) and finishing deep in the anodic region (+6 V). In the cathodic region, negative currents were experienced with bubbles observed to form readily on the surface of the working electrode. Linear sweep curves subsequently displayed an initial plateau, -1 to 0.5 V, producing no current. Further into the anodic region, a short exponential rise, region 1, occurred before an extended linear rise, region 2. Curves subsequently experienced a turning point, region 3. Between +3 V and +5 V, the current falls to a plateau, regions 4 and 5, limiting the current density to ~25 mA/cm$^2$. The plateau region is sustained to at least +10 V.

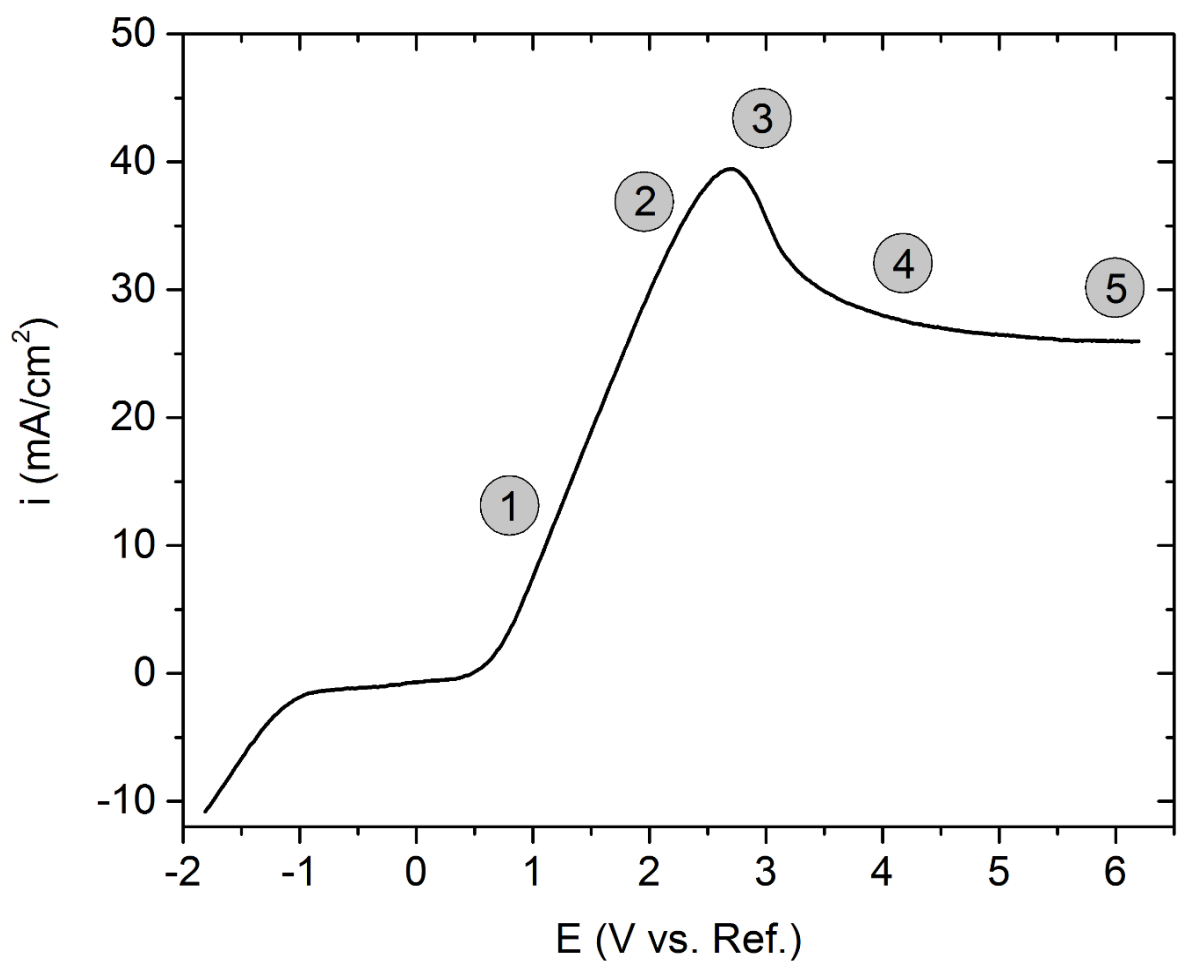

*Figure 2: Typical linear sweep curve showing the potentials used in the potential study labelled above the curve. Data taken for the LS curve was obtained using a scan rate of 50mV/s.*

It would be expected that the current rises again as oxygen is evolved, however high acidity and an appreciable lack of water (and therefore OH$^-$ ions) in the solution is thought to be precluding this reaction, pushing it to potentials greater than the capabilities of the potentiostat used.

Increasing scan rates were found to increase the total current generated as expected; higher scan rates produce a smaller diffusion layer permitting greater current. Additionally, the peak was observed to increase to higher potentials for higher scan rates indicating slow kinetics with respect to the experiment and irreversible reactions.

Cyclic voltammetry was performed to further examine the kinematics and electrochemical stability of the system. Between the first and second cycles, the current increases for anodic potentials. This can be attributed to the time delay in the solution finding a steady state. Cyclic voltammetry showed the polishing reaction to be irreversible with no affinity for the polished uranium species to redeposit on the working electrode once the potential has been swept back to the cathodic region. Instead, this region was dominated by the evolution of gas, likely to be $H_2$.

### Chronoamperometry

A typical chronoamperometry curve is shown by the black line in Figure 3. The same data is plotted against the square root of time (Cottrell plot) and on a logarithmic axis shown by the red and blue curves respectively. Initially, the current is high, close to 100 mA/cm$^2$. This regime continues until ~0.1s, region 1, before the current decreases much more rapidly, achieving a new steady state that exists from between 1 and 20 seconds, region 2. The logarithmic curve, blue line, suggests a final regime which seems to only be emerging at times of beyond 200s.

Analysis of the gradients of the current on the Cottrell plot, red line, allows for assessment of the diffusion coefficients associated with each regime. Due to the relatively high uncertainty of some of the coefficients of Cottrell's equation [13], a better measure is the ratio between the two. The diffusion coefficient in region 1 was found to be $(1.02 \pm 0.32) \times 10^7$ times larger than that of region 2. It is thought that region 1 represents a transfer of charge through the layer initially adjacent to the working electrode surface, whereas region 2 relies on the diffusion of ions to the vicinity of the sample to continue the production of current. Region 3 could be arising from changing chemistry either in the bulk solution or due to the growth of a very thin oxide film.

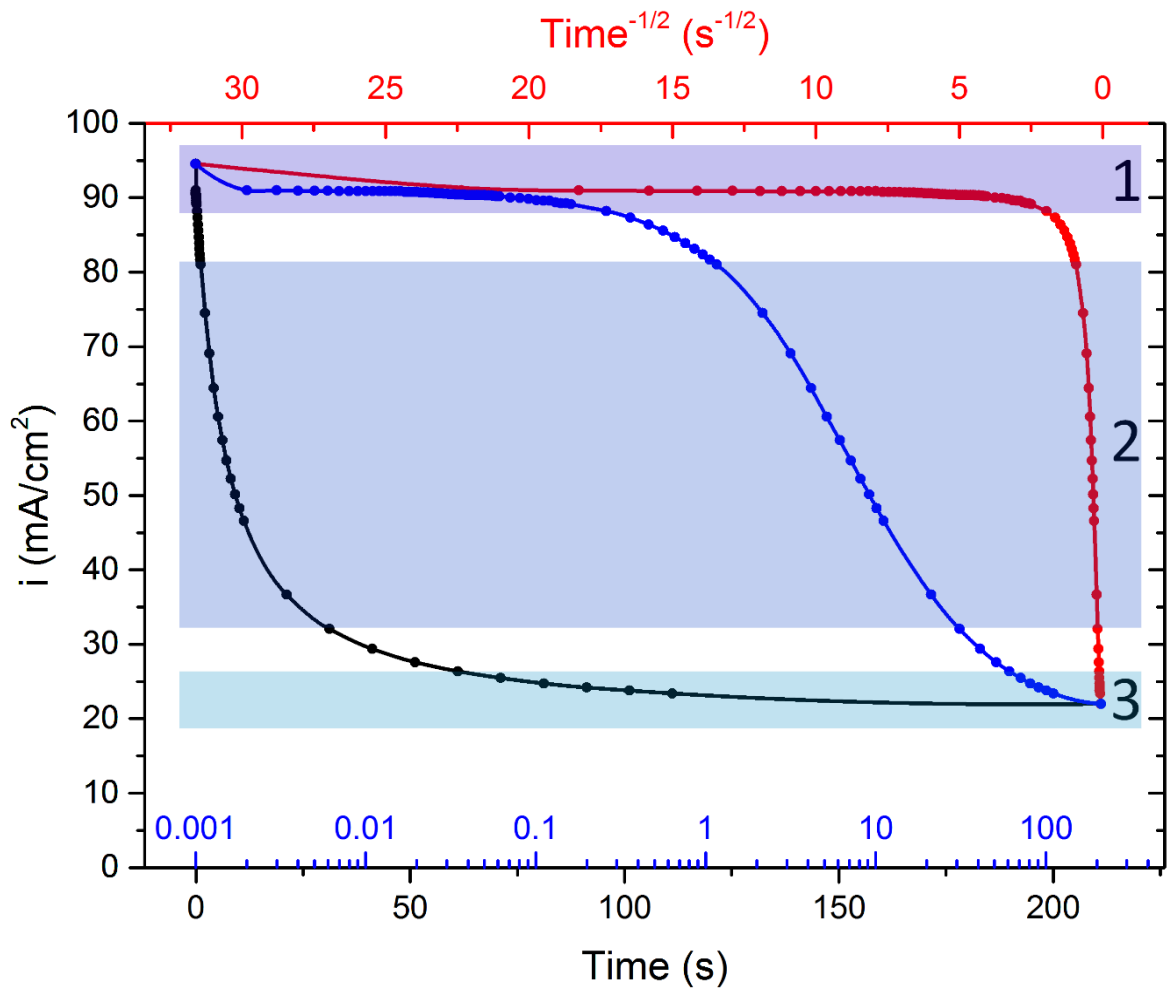

*Figure 3: Chronoamperometry plot showing the behaviour of the current with respect to time. The black curve is plotted on a linear axis, whereas the red curve shows a conventional Cottrell plot and the blue curve has been plotted with a logarithmic x axis. Each of these curves have been plotted to show the passage of time passing from left to right.*

## Polishing Potentials

To confirm the optimal polishing potential, five samples cut from the same batch were examined at varying potentials corresponding to unique points on the linear sweep curve, Figure 2. Polishing time was kept fixed for all samples in this study at 210 seconds. The resultant EBSD maps are shown in Figure 4. Average confidence index was observed to rise with increasing potential with a maximum at 4 V.

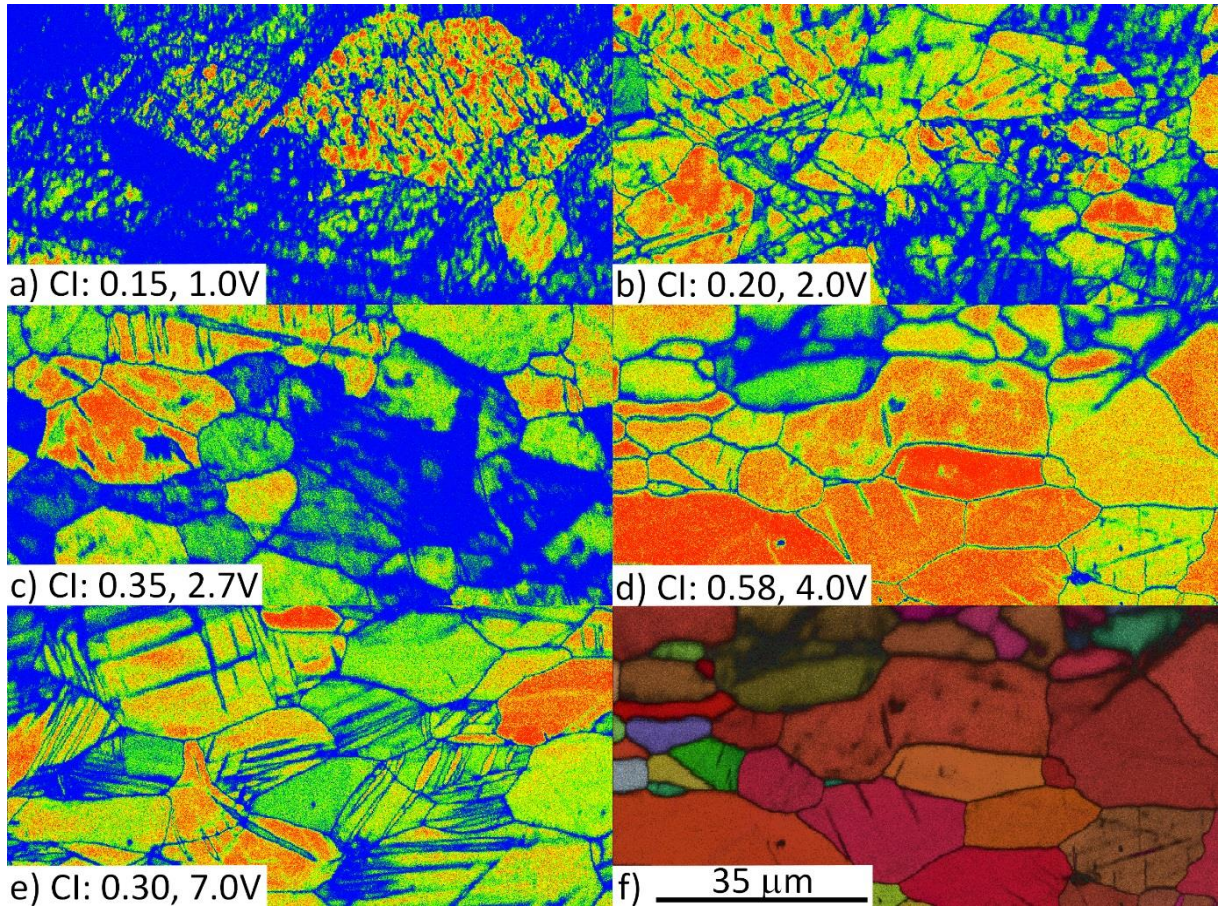

*Figure 4: Images a through e are EBSD maps of samples which had experienced progressive polishing potentials (all for 210 s) showing confidence on the colour-scale. Image f shows a map of the best polished sample, image d, with contrast derived from confidence index and the colour from crystallographic orientation. Scale bar serves all maps which are 100 x 50 µm in size.*

Figure 4 appears to agree well with basic electropolishing theory [14]. Whilst the potential is less than the peak of the linear sweep curve, the polishing is poor and EBSD patterning is largely unsuccessful. Pore formation was additionally observed at low voltages which may have been contributing to poor patterns through a non-flat surface. It is assumed at these low potentials, the amount of material removed has been insufficient to allow sampling of material free from mechanical damage. As the potential rises and enters the plateau region, the polishing is more effective and confidence indices increase. Given a relatively short polishing time and originally rough surface, samples have quite varied CI's at this stage. There may also be issues with sample homogeneity. However, the 4.0 V and 7.0 V samples are indexing the entire sample effectively, and a potential in this range has proven to be effective in producing good EBSD maps. Chronoamperometry scans of the 5 test samples are shown in Figure 5. All potentials representing stable polishing were ultimately limited in current density at 20-25 mA/cm$^2$ at 210s. 30 mA/cm$^2$ has been reported as optimal polishing conditions using a similar solution [15].

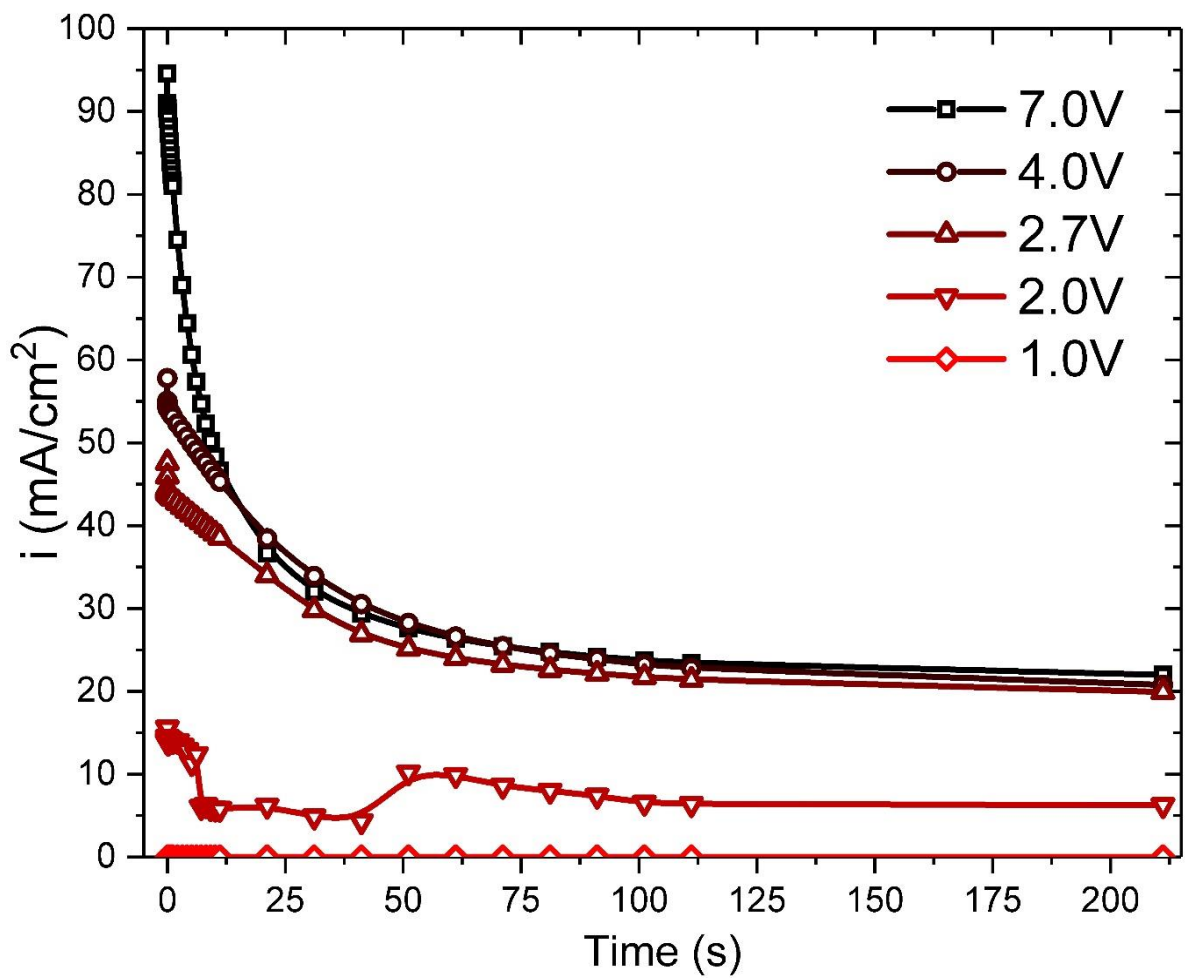

*Figure 5: Chronoamperometry scans pertaining to the 5 samples. As the potential of the chronoamperometry step is increased, the instantaneous current also increases. For higher potentials, steady-state current is limited at roughly 23 mA/cm$^2$. Potentials of 1.0 or 2.0 V were insufficient in producing reliable polishing. Data points have been joined up with spline curves to simply show the trends.*

## Polishing Durations

Total polishing time was subsequently examined by varying polishing durations whilst maintaining the potential at 5.5 V, as was found to be within the suitable range of potentials in the previous section. Maps received from this study are shown in Figure 6.

Though there appear to be slight differences in the microstructure, such as the prevalence for twinning, stronger patterns were achieved for longer polishing durations. Alternatively, deformation twinning may have been induced by mechanical preparation, but polishing has removed sufficient material to observe mostly un-deformed material, as shown in images e and f. In addition to the polishing duration, it appears that grain orientation is strongly affecting the pattern certainty. The pair of images, e and f, show many grains have uniform confidence indices within a grain whereas their neighbours have a lower confidence that is also uniform within the grain. Reasons for this result may be due to the preference that a grain may have for electropolishing due to superior polishing rates [16], initial inherent oxide thickness or the number of lattice planes contributing to the scattering of electrons in each crystallographic orientation. There is also another effect relating to the density of Kikuchi bands crossing the detector screen given a grain's orientation.

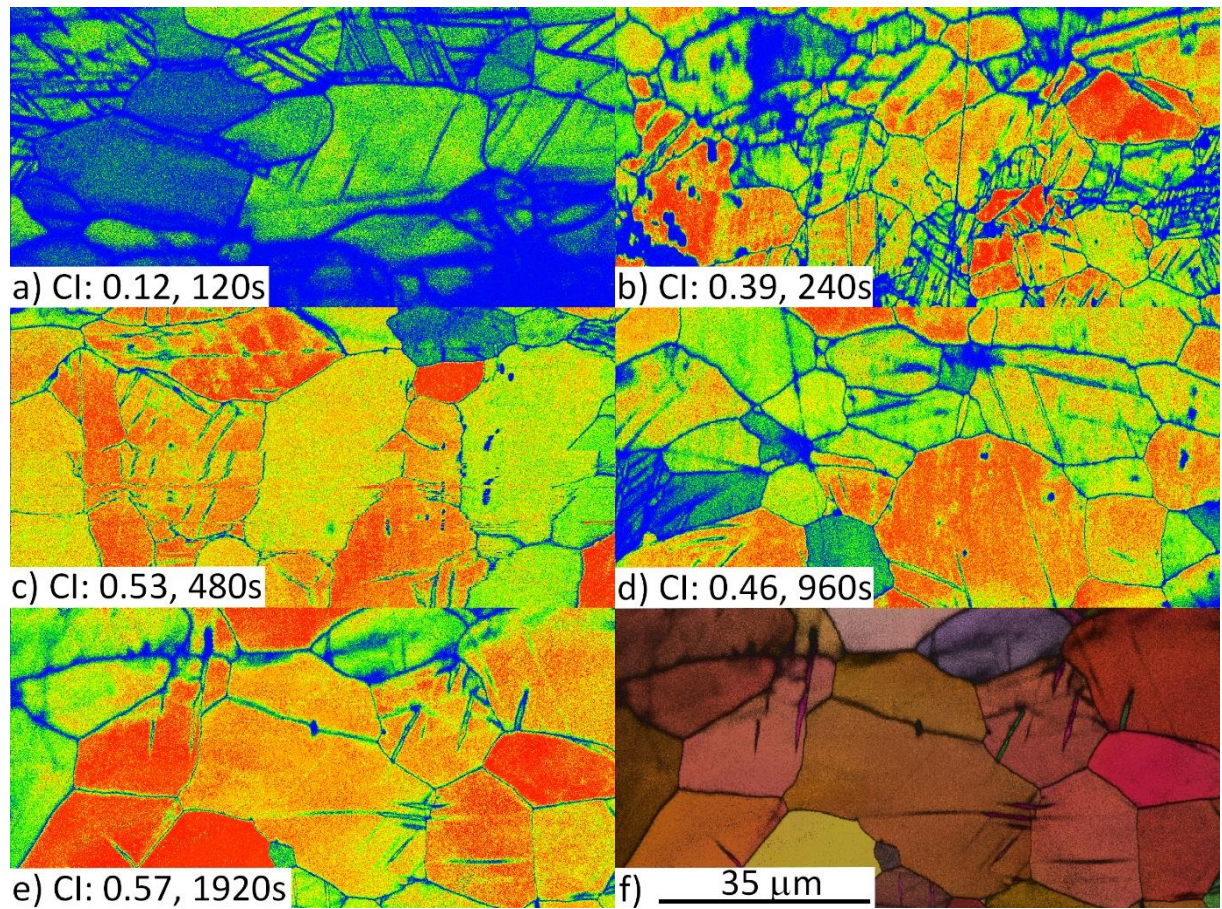

*Figure 6: Images a through e are EBSD maps of samples which had experienced progressive polishing durations (all at 5.5 V) showing confidence on the colour-scale. Image f shows a map of the best polished sample, image e, with contrast derived from confidence index and the colour from crystallographic orientation. Scale bar serves all maps which are 100 x 50 µm in size. Beam drifted in image c resulting in a disjointed image.*

Some papers have found that lengthy polishing time has the potential to roughen the surface after a point, subsequently reducing the pattern quality [17]. In this case, the polishing is occurring on a sufficiently slow scale and as the material possesses a single phase, this effect does not appear to be contributing. The surface has been made considerably smoother following the final mechanical polishing stage to enable the production of good EBSD patterns.

## Evaluation of Residual Stresses

A final sample was prepared to measure residual stresses and record EBSD maps. Mechanical polishing was extended to 3 and 1 μm diamond pastes to ensure the surface was scratch free. The sample was subsequently electropolished for 600 s at 5.5 V. Figure 7 shows an EBSD map of many grains including a clustering of carbide inclusions. Although most of the carbides have been stripped out in the polishing process, evidence of crystal structure remains in their pits and is capable of being indexed. Due to the shape of the carbide pits, only portions are flat and shallow enough to direct the electron beam to the detector to be indexed.

The reason for the preferential stripping of the carbides is unclear. With a lower conductivity than the metal, it might initially be assumed that the metal should be preferentially etched, 'popping-out' the carbides. Alternatively, the heightened hardness of the carbides leads them to protrude the surface following mechanical polishing, causing them to disrupt the Helmholtz double layer and focus polishing in the vicinity. Pockets of higher contrast in Figure 7 show regions where the confidence index was higher around ejected carbides. In these regions, quicker polishing of metal may have caused material behind the carbide to be polished much more quickly until they 'pop-out'.

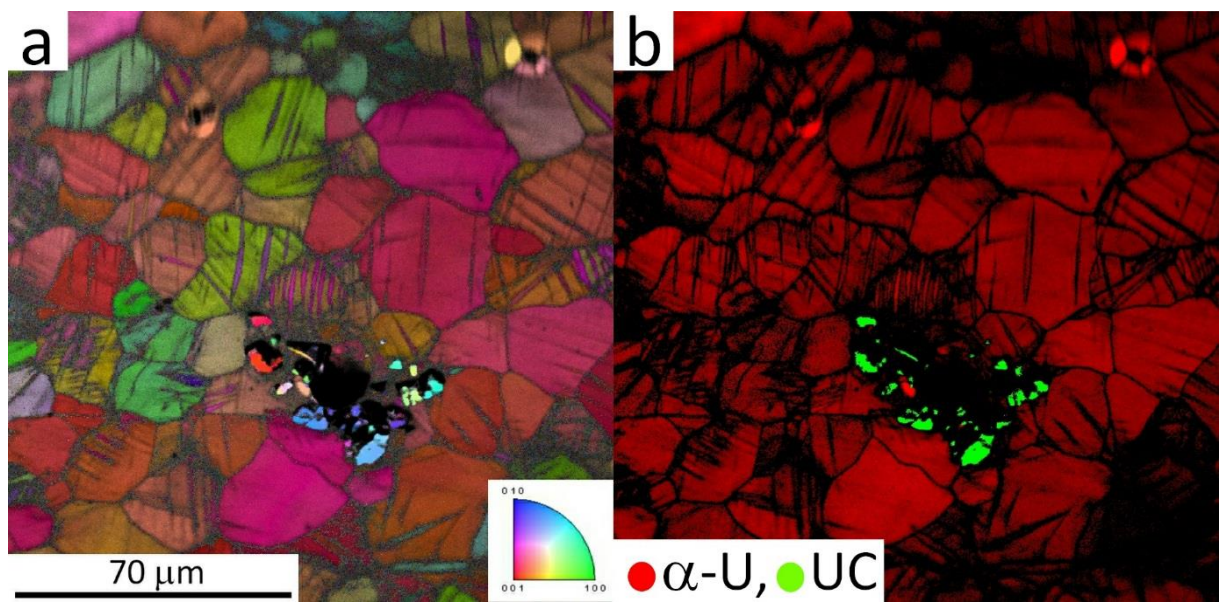

*Figure 7: a) EBSD CI+IPF map of a pristinely finished sample. A large proportion of the sample was indexed in the high-resolution map with primarily grain and twin boundaries responsible for regions un-indexed. In the polishing process, in this case, a clustering of carbides was preferentially removed. In the resultant pit, portions were capable of being mapped using a UC material file. b) shows the distribution of the two phases.*

In a high-resolution scan, Figure 7, substantial deformation to crystallites may be observed. The (130) twinning mode is particularly prevalent in the material matching literature observations [5], with kinks also seeming to appear possibly due to slip. Most of the grains probed in this scan have returned a reasonable to good EBSD pattern leading to a broadly gaussian confidence index distribution with a mean of 0.36 and a standard deviation of 0.22. The effect of grain orientation is thought to be contributing to the substantial variation in confidence indices.

Large EBSD maps were also acquired from the same sample as shown in Figure 8. Due to the larger step size required for this scale (covering over 1 mm x 0.5 mm), twins are not easily observed. The preferred orientation is noticeable in this map and the modelled inverse pole figure texture plot is inset. Due to the mechanical working of the material during the manufacturing process, grains are biased towards the [001] direction oriented normal to the surface.

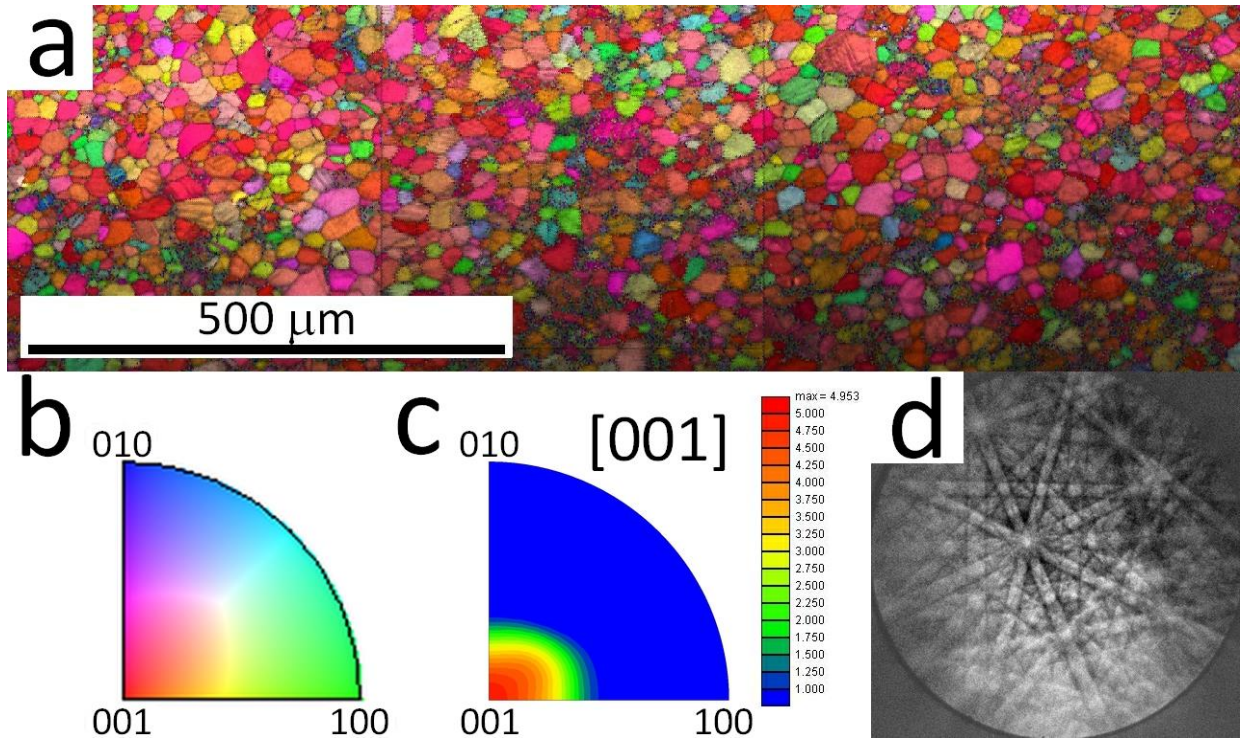

*Figure 8: a) Large EBSD map measuring 1.5 x 0.5 mm. Due to the large map, the effect of sample tilt and low magnification acts to reduce quality of focus (and therefore pattern quality) at the bottom of the image. b) Inverse Pole Figure map orientation legend. c) Unit stereographic triangle of the [001] oriented inverse pole figure texture plot. d) Typical Kikuchi pattern obtained from the final sample in focus. 1x1 binning was implemented with no background subtraction for this pattern to assess feasibility of cross correlation EBSD.*

X-ray diffraction data from the optimised sample is shown in Figure 9 with a Rietveld refinement fitted powder profile. Crystallographic texture is clearly prevalent in this material when compared to a standard powder pattern. Intensities are much higher for (002) and (112) reflections and diminished for the (021) and (131) planes. This result agrees with the Inverse Pole Figure obtained via EBSD analysis, inset of Figure 8. The Pole Figure generated through XRD strongly correlates with straight rolled foils examined by Einhorn et al. [18].

## Williamson-Hall Method

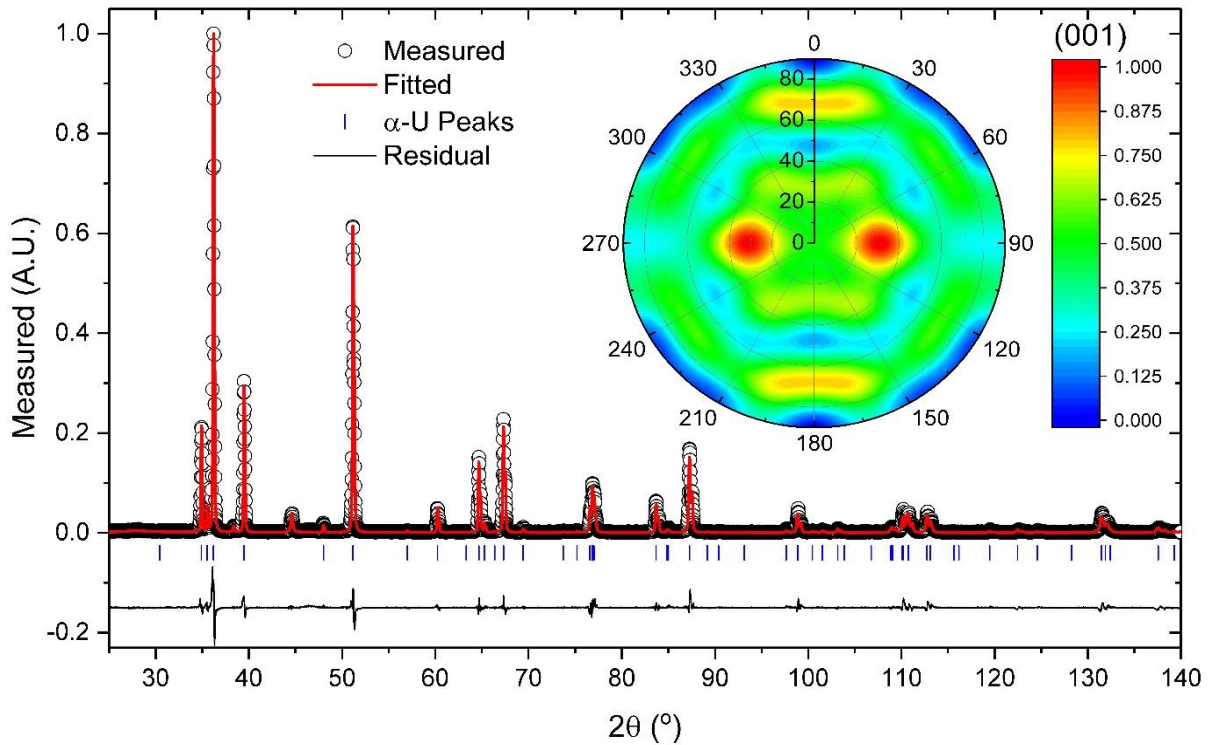

*Figure 9: Specular XRD scan fitted using Rietveld refinement. Inset shows the result of fitting the texture using spherical harmonics.*

Williamson-Hall peak analysis was performed on the specular powder pattern to assess the material for strain. Peak widths were observed to increase at higher diffraction angles, as shown in Figure 10. This small increase observed, <1 x 10$^{-3}$, over such a large q range illustrates that there is little deviation in the periodic structure within the sampled volume. Using the formula,

$$\beta_{hkl} \cos\theta = \frac{K\lambda}{D} + \frac{\sigma_{hkl} \sin\theta}{E_{hkl}},$$

(2)

where β is the peak width, D is the particle size, K is Scherrer's constant and $E_{hkl}$ is the in-plane elastic modulus, the average strain, Kλ/D, and stress of each measured reflection, $\sigma_{hkl}$, can be assessed. In an orthorhombic system, such as uranium, $E_{hkl}$ is given by,

$$(E_{hkl})^{-1} = n_1^4 s_{11} + n_2^4 s_{22} + n_3^4 s_{33} + n_2^2 n_3^2 s_{44} + n_1^2 n_3^2 s_{55} + n_1^2 n_2^2 s_{66} + 2n_1^2 n_2^2 s_{12} + 2n_1^2 n_3^2 s_{13} + 2n_1^2 n_3^2 s_{23},$$

(3)

where $n_i$ are the Euler angles between reflections and the principal axes, $s_{ij}$ are the inverse of the elastic stiffness moduli, $c_{ij}$, ($s_{ij}=c_{ij}^{-1}$) obtained from Fisher and McSkimin [19]. Using the in-plane elastic modulus produced a stress of -83 ± 33 MPa. In comparison, treating the material as having undergone uniform compressive stresses and possessing an elastic modulus of 201 GPa [20], a crude assessment of the stress may be evaluated as -45 ± 9 MPa. The reduced scatter of the fitted line and the resulting lower error suggests that this is the better model. Rather than behaving in a way that might be expected of it given sufficient degrees of freedom, the material has deformed uniformly normal to the surface. The various deformation pathways of uranium, such as twinning and slip, have not been accounted for using this method.

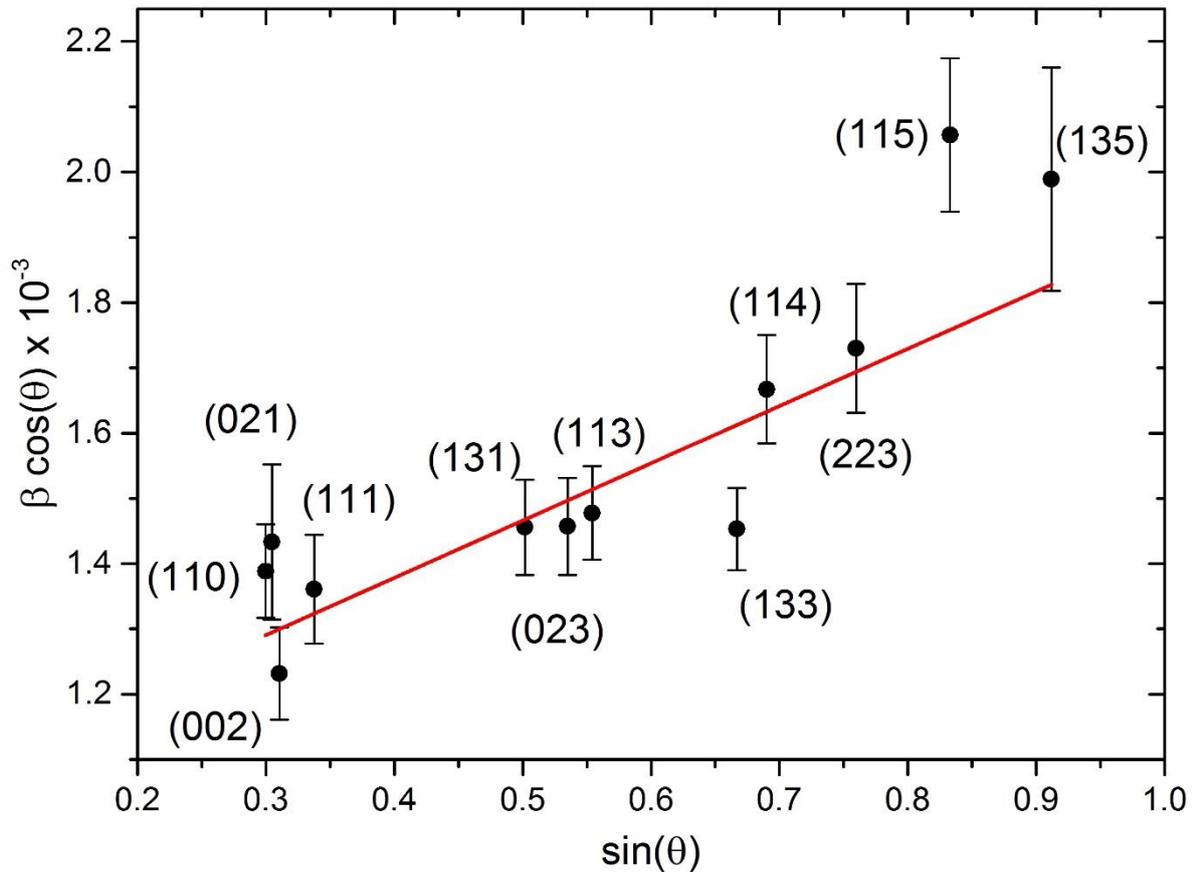

Figure 10: Williamson-Hall plot of fitted distinct uranium peaks from Figure 9. The red line denotes a linear best fit with gradient of (8.8 ± 1.9) x $10^{-4}$. A couple of outliers might indicate that the complex mechanical deformation pathways of uranium is affecting some planes greater than others.

### Sin²ψ Method

Small shifts in peak positions were observed on the order of thousandths of Angstroms. For the high values of 2θ used, this approaches the resolution of the diffractometer. Peak shifts to higher values in 2θ, indicate decreasing interplanar distances arising from compressive stresses. This fits with the expectation that mechanical working compresses most crystal orientations.

Given the results from the last section, the bulk elastic modulus was used to determine the stress for each reflection using,

$$\sigma_{hkl} = \frac{E}{1+v} \frac{d\varepsilon}{d(sin^2\psi)},$$

(4)

where $v$ is Poisson's ratio for uranium, 0.23 [20], and ε is the strain between the ideal and measured peak position, based on a Rietveld fitting of the entire pattern. The stress can therefore be extracted from the gradients of the curves in Figure 11.

Figure 11 shows the relationship between interplanar spacing and sin²ψ. Under uniform stress, the relationship should be linear. In this experiment, curves appear to be well modelled by a parabolic curve. The stress profile is not constant but increases closer to the surface. To assess the stresses for each crystallographic direction, the fitted curves were differentiated analytically and evaluated at the highest available value of sin²ψ. This gave an upper estimate of the stress for the material closest to the surface. Using the uniform stress deformation model, stresses were determined to be $\sigma_{223}$ = -448 ± 227 MPa, $\sigma_{115}$ = -841 ± 156 MPa and $\sigma_{135}$ = -815 ± 168 MPa. The stress evaluated at sin²ψ=0, where the same volume is sampled under the Williamson-Hall method, produces a consistent result (averaged -44 ± 44 MPa for the (115) and (135) reflections). As the compressive yield strength of uranium is roughly 750 MPa [5], only the top few layers could be slightly plastically compressed.

The surface value is substantially higher than as evaluated by the Williamson-Hall method, as the depths profiled by this technique are far shallower. Small changes, as progressively less volume is sampled, illustrates that the effect of mechanical polishing has been greatly reduced. After a reasonably shallow depth, the material is akin to that of the bulk possessing residual strain incurred during manufacture. It is expected that further polishing could be used to completely reduce all artificial stress in the material.

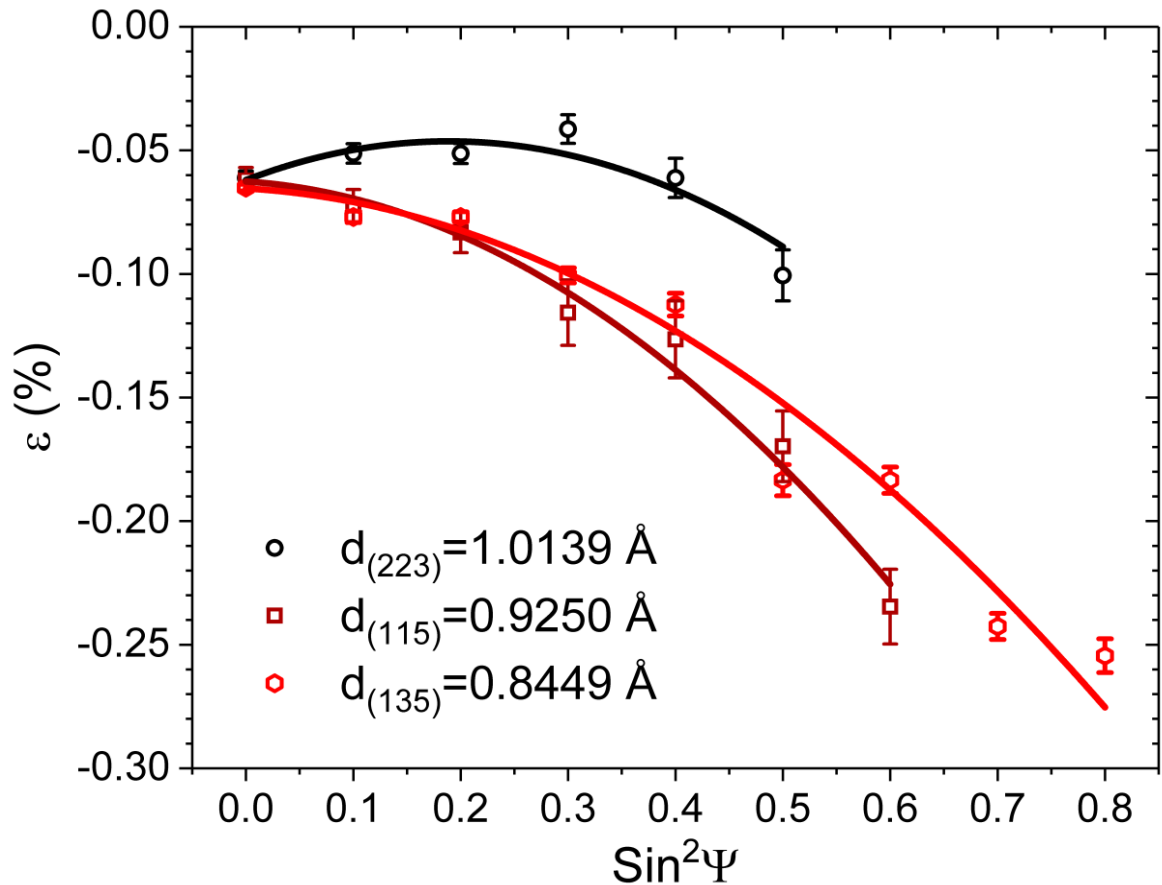

Figure 11: Sin²ψ plot for 3 of the more intense reflections at the higher end of uranium's XRD powder pattern. Reflections relating to smaller interatomic spacings permit the greatest tilting in the ω-axis mode, thereby allowing shallower measurements, see Equation 1.

## Discussion

Using a systematic approach, an appropriate polishing potential of at least 3 V has been found to be sufficient for this system. Potentials significantly less than 3 V has been shown to produce unsatisfactory results. This is possibly due to an inability to form a sufficiently strong double layer promoting consistent polishing. Whilst there was little difference above 3 V in terms of quality of EBSD patterns received, it is the authors' belief that oxide formation should be more prevalent over time and operating at higher voltages shows no gain in terms of quality of polish so should be discouraged. This work has highlighted that there is plenty of scope to expand, particularly using different solutions to gauge the effectiveness of each.

XRD investigations confirm the lack of oxide present on the sample and that the metal is highly textured. Peak profile analysis from the specular data using the Williamson-Hall method shows little micro-strain throughout the metal. This is reinforced by the Sin²ψ data illustrating that small strains and stresses are observed but increase towards the surface of the material. These two experiments show that the mechanical and electrical polishing undertaken has produced a low-strain surface, representative of bulk material.

This low-stress surface is also evident through the quality of EBSD patterns measured from uranium, a high Z-material. Large areas accessible using this polishing procedure illustrate an ability to acquire information on the scale of millimetres, crucial for material characterisation and many real-world corrosion experiments. Mapping such large scales has allowed for the assessment of useful material characteristics such as grain size and texture.

Kikuchi patterns obtained in this study have enabled the interrogation of twin and carbide formation. The authors believe that the surface and pattern quality produced in uranium from this optimised process are of a sufficiently high quality for Cross-Correlation EBSD, an example of a typical EBSD pattern is shown in Figure 8. This would allow investigation of the origin of stresses within the material, which are likely to be focused around crystallographic defects such as twins and grain boundaries.

The influence of grain orientation is a large factor in the correct assignment of phases and could be assisted by further investigation. Given the low crystal symmetry, it would be reasonable to expect differences in image quality and therefore confidence index to arise from crystallographic orientation [21]. Since stresses are low and there is no obvious oxide present, the effect is expected to be due to crystallographic orientation.

## Conclusions

This paper has illustrated in that in uranium, production and characterisation of low-stress surfaces is possible using a relatively simple setup. In this case, optimal polishing potentials were found to occur beyond 3 V, with 5-10 mins sufficient to produce good patterns from an initially quite rough surface. The work has shown development of the processes of electropolishing, enabling a standard protocol for the assessment of electropolishing procedures. Ultimately, a combination of mechanical and electrochemical surface preparation has been used to produce very uniform regions allowing for the reliable mapping of large surfaces using EBSD. It is key to illustrate that the resultant surface and strain state produced is not a result of the preparation method, but representative of bulk material. Therefore, this method can be utilised as a tool for interrogating microstructures as well as acting as a model start point for corrosion or oxidation experiments.